\documentstyle[preprint,aps]{revtex}
\begin{document}
\draft
\title{ The Interest in Kaonic-Deuterium Atoms }
\author{S. Wycech\thanks{e-mail "wycech@fuw.edu.pl"}}
\address{Soltan Institute for Nuclear Studies,
Warsaw, Poland}

\maketitle
\begin{abstract}

K meson-deuteron scattering lengths, atomic level shifts and widths
are calculated. These quantities depend on controversial 
kaon-nucleon scattering amplitudes extrapolated by few MeV  
below the $KN$ threshold.  Experimental consequences of several  
$KN$ interaction models are studied.
\end{abstract}

\newpage

\section{Introduction}
\label{intro}

Few-body interactions of the $K^{-}$ meson may complement our 
knowledge on the kaon-nucleon interaction. One point of interest 
in this field is the old question of the nature of $\Lambda(1405)$
state interpreted as an elementary particle \cite{dali} or a $KN$ 
quasi-bound state\cite{amar}. This uncertainty is aggravated by a 
serious discrepancy between the value of  $Kp$ scattering length 
obtained in scattering experiments \cite{cibo} and the value extracted 
from the  $1S$ level shift in kaonic-hydrogen atoms 
\cite{davi},\cite{izyc},\cite{bird},\cite{iwas}. 
Of related interest is the possibility of $K$-nucleus quasi-bound 
states \cite{wyce}, the question of kaon condensation and its 
interesting astrophysical consequences \cite{brow}. Another 
controversial, misunderstood 
but appearently related result is the anomalously 
large $2P$-state width and shift detected in the $K^{-}$He 
atoms \cite{wieg},\cite{bair},\cite{battbs}. The helium 
measurements indicate possible  existence of a nuclear 
bound state in the $K^{-} \alpha$ system. 
    
These unsolved problems have been dormant for a decade, due 
to lack of intense low energy kaonic beams. Such a beam is now coming 
into operation at the DA$\Phi$NE facility and the kaonic atom  
program DEAR is being installed there \cite{dear}. 
Another kaonic atom program is running now at KEK \cite{iwas}.
The purpose of this letter is to study the physical interest 
and chances of  the 
kaonic deuterium X-ray measurements.

The $K^{-}$ deuteron system is in general difficult to 
describe because of numerous inelastic decay channels. 
However, to study 
the kaonic-deuterium atoms one needs  to know only 
the $KD$  elastic scattering amplitude at threshold. 
This could be obtained from the elastic $KN$ amplitudes by 
summation of the multiple scattering series. Such a description 
involves elementary  $KN$ scattering amplitudes extrapolated 
into  the subthreshold region. Such extrapolation is necessary 
since the nucleons are bound. In addition, the recoil 
energy of the $KN$ pair reduces the energy available in its
C.M. system by several MeV. Thus in the $KD$ atoms 
the energies allowed in the $KN$ subsystems are located 
half-way between the $KN$ threshold and the $\Lambda(1405)$. 
On the other side, the $K^{-}$He atoms involve $KN$ energies 
down in the interesting $\Lambda(1405)$ region as the nucleons 
are bound more strongly in the $\alpha$ particle. The price to 
pay is the difficulty in a reliable description of the $K \alpha$ 
system. It is the deuteron that offers a good compromise between 
the requirements of theories and the atomic experiments.

In this letter a simple formula for the $KD$ 
scattering length, volume etc. is given  and related shifts 
and widths of atomic levels are calculated. This formula is 
expressed by certain averages of the $Kn$ and $Kp$ $S$-wave 
scattering amplitudes below the thresholds. These averages 
are our free parameters and are taken from
theoretical models or phenomenological scattering analyses. 
It turns out that it is the widths of atomic states that are 
most sensitive to these imput parameters. The $1S$ level shifts 
display rather limited sensitivity to the real amplitudes 
and are dominated by the absorptive parts of $KN$ amplitudes. 
The results indicate a good chance for a successfull measurement 
within the DEAR project.

\section{Relation of level shifts to scattering amplitudes} 
\setcounter{equation}{0}

The atomic level shifts $\Delta E $ induced by nuclear 
interactions may be related to the scattering lengths of 
the orbital meson on atomic nuclei. These relations 
are very simple  when the 
atomic Bohr radius $B$ is larger than 
the lengths  characteristic for nuclear interactions : 
the nuclear radius $R_D$ and the meson-nucleus scattering 
length $A_{0}$.  For the $S$ waves such a relation  

\begin{equation}
\label{C1} 
\Delta E_{nS}= E_{nS}-\epsilon_{nS}= 
\frac{2\pi}{\mu}|\psi_{n}(0)|^2A_{0}(1- \lambda A_{0}/B) 
\end{equation}
is known as the Trueman formula \cite{true}. It is accurate to a 
second order in $A_{0}/B$, good enough for the 
$KD$ system. The electromagnetic energy $ \epsilon_{nS}$ in 
Eq.({\ref{C1}) is composed of the Bohr's atom energy  
$ \epsilon_{nS}^{o}= -\mu(\alpha)^2/2n^2$ corrected for 
relativity and effects of the deuteron electric polarisability. 
The Bohr radius is given by $=1/\alpha \mu $ where $\mu$ is the 
reduced mass. 
In the $1S$ state one has $\lambda =3.154$,$ B=69.2fm$, $ A \approx 1fm$  
and the second order term in Eq.({\ref{C1}) constitutes a few 
percent correction. Such corrections are negligible in higher 
angular momentum states. For these states a simpler linear relation 

\begin{equation}
\label{C2} 
\Delta E_{nL}= \epsilon_{nL}^{o} \frac{4}{n} \Pi_{i=1}^{L}
( \frac{1}{i^2}-\frac{1}{n^2} )A_{L}/B^{2L+1}
	     = - \Theta_{nL} \frac{ A_{L}}{\mu B^{2L+1}}
\end{equation}
due to Lambert \cite{lamb} is sufficient. We shall be interested in
the lowest circular states $ 1S,2P,3D $. For these  
the overlap factors  $\Theta $ equal correspondingly 
$2,3/16,60/(81)^2 $. 

Our problem is to calculate the $KD$ scattering length $A_{0}$,
scattering volume $A_{1}$ etc. These quantities are 
obtained in the next section by a trick which sums the $KD$ 
multiple scattering series. This method was tested with success 
in the $\eta D$ calculations  where the details of 
the method are explained  \cite{etad} .  In this way the kaon-deuteron 
scattering lengths are expressed in terms of deuteron 
radius, properties of $NN$ interactions  and the 
$KN$ scattering amplitudes. The latter are complex 
due to $KN$ coupling with the open  $\Sigma \pi $, 
$\Lambda \pi $ channels. The
absorptive parts af these amplitudes contribute to absorptive parts 
of the $KD$ scattering lengths. The latter determine the
widths of atomic $KD$ states.

\section{A Formula for the kaon-deuteron scattering matrix}
\label{sec2}

The purpose of this section is to exploit a simple formula
that relates the meson-deuteron scattering lengths  to the
meson-nucleon subthreshold scattering amplitudes. 
The $KD$ lengths are found by a summation
of the multiple scattering series and are expressed in terms 
of few basic multiple scattering integrals. 
Although we are interested in all partial waves in the $KD$ system 
only the $S$ wave interactions in the $KN$ system are considered. 
To keep the presentation 
simple, the impulse approximation is discussed in some detail 
but higher order terms and the summation of the series is 
presented rather schematicaly. For details we refer to 
a similar calculation of $\eta$-deuteron scattering length 
\cite{etad}.

Let us begin with a multiple scattering expansion that follows 
from the three-body Faddeev equations for a meson interacting 
with a pair of nucleons labeled 1,2. For the situation of 
meson-deuteron scattering below the deuteron breakup, the series 
for the kaon-deuteron $T_{KD}$-matrix is:
\begin{eqnarray}
\label{f2}
T_{KD} & = & t_1+t_2+t_1G_0t_2+t_2G_0t_1+t_2G_0t_1G_0t_2+t_1G_0t_2G_0t_1 \nonumber \\
& + & (t_1+t_2)G_{NN}(t_1+t_2)+(t_1+t_2)G_{NN}(t_1+t_2)G_{NN}
(t_1+t_2)+...
\end{eqnarray}
where $t_i$ are meson-nucleon  scattering matrices, $G_0$ is
the free three-body propagator and $G_{NN} = G_0 T_{NN} G_0$
is that part of
the three-body propagator which contains the nucleon-nucleon
scattering matrix $T_{NN}$. This expansion is
performed in the momentum space and appropriate integrations
over the intermediate momenta  are understood.

The scattering amplitude is determined by an average
\begin{equation}
\label{f3} 
<\psi_D \psi_{K}^L \mid T_{K D}(E) \mid \psi_D \psi_{K}^L>
\equiv <L \mid T_{K D}(E) \mid L>
\end{equation}
where $\psi_D$  is the deuteron and $\psi_{K}^L$ is the
free-meson wave function for an $L$-th  partial wave  
in the meson-deuteron system. We define the $K D$ 
scattering lengths as the limit 
for $p \rightarrow 0$ of the expression 
\begin{equation} 
\label{f4} 
A_{K D}(L) = -(2\pi)^2 m_{K D} < L\mid T_{K D}/p^{2L} \mid L > 
\end{equation}
where $p$ is a relative meson-deuteron momentum and $m_{K D}$ is 
the corresponding reduced mass.  
The normalization in Eq.\ref{f4} is related to the normalisation 
of  $t_{K N}$  
\begin{equation}
\label{f10} 
t_{K N}(E)= - \frac{a_{K N}(E)}{(2\pi)^2 \mu_{K N}}
\end{equation}
where $a_{K N}(E)$ is the $S$ wave scattering amplitude
in the $KN$ system normalised to the scattering length at 
the $KN$ threshold. Here, $\mu_{K N}$ is another reduced mass.

Calculations of the multiple scattering integrals are
straightforward although tedious and the formulae are lengthy. 
For an illustration we reproduce the leading impulse approximation
lengths for zero range $KN$ forces. Actual calculations are
performed with a finite force range generated by a separable 
Yamaguchi potential of 600 MeV inverse range .

The leading term 
$<L\mid T^0_{K D}\mid L > = <L \mid t_1 + t_2 \mid L >$ 
is the impulse approximation. This 
term yields  the impulse  $KD$ scattering lengths 

\begin{equation}
\label{f11} 
A_{KD}^0(L) = \frac{\mu_{KD}}{\mu_{KN}}(\bar{a}_{Kp}+\bar{a}_{Kn})
	     \frac{<(r/2)^{2L}>_D }{(2L+1)!!)^2}
\end{equation}

where $<(r/2)^{2L}>_D$ is the $2L$-th radial moment  of the deuteron 
related to its center  and 
\begin{equation}
\label{f12} 
\bar{a}=\int a(-E_D -\frac{p^2}{2m_{N,K N}})\mid
\tilde{\phi}_D^L(p)\mid^2 d\vec{p} 
 \int \mid \tilde{\phi}_D^L(p)\mid^2 d\vec{p}
\end{equation}
are the scattering matrices  averaged over some energy region, generated 
by the recoil of the spectator nucleon. The range of the latter is 
given by a Bessel transform of the deuteron wave function 
\begin{equation}
\label{f13} 
 \tilde{\phi}(p) = \int \psi_D(r)j_L(pr/2) p^2 dp  
\end{equation}

The energies involved in Eq.(\ref{f12}) extend down to
the unphysical subthreshold region. The recoil energy region 
given  by the Bessel transform depends on the partial wave $L$. 
The relevant distributions peak around energies of 
-12,-7 and -5 MeV for $ L = 0,1,2 $ correspondingly. 
Therefore, a model is required for this extrapolation and
several possibilites are discussed later.

Now a partial summation of the series (\ref{f2}) for $<T>$ 
is obtained by
\begin{equation}
\label{f5} 
<T^1_{K D}>=\frac{<T^0_{K D}>}{1-\Omega_1-\Sigma_1}
\end{equation}
with 
\begin{equation}
\label{f6} 
\Omega_1 = \frac{<t_1G_0t_2+t_2G_0t_1>}{<T^0_{K D}>} 
\end{equation} 
\begin{equation}
\label{f6a} 
 \Sigma_1  =  \frac{<T^0_{K D}G_{NN}T^0_{K D}>}{<T^0_{K D}>} .
\end{equation}
This partial sum is very close to the formula for meson 
scattering on two fixed nucleons \cite{brue}. It contains
terms of the order $\xi=<t>/R_D$ in the denominator. The 
parameter $\xi$ does not need to be small for the success of
Eq.(\ref{f5}), which works also for $|\xi| > 1$, when the multiple
scattering is divergent.
In the $K-D$ case $|\xi|$ falls into 0.3-.5 range.
Corrections of higher orders of $\xi$  in the denominator of
Eq.(\ref{f5}) may be obtained by comparing higher orders in Eq.(\ref{f2})
with a series expansion of Eq.(\ref{f5}) with respect to $\Omega_1$ 
and $\Sigma_1$. In this way the second approximation is obtained
\begin{equation}
\label{f7} 
<T^2_{K D}>=\frac{<T^0_{K D}>}{1-\Omega_1-\Sigma_1
[\Omega_2-(\Omega_1)^2]-[\Sigma_2-(\Sigma_1)^2]-[\Delta_2-\Omega_1\Sigma_1]}
\end{equation}
where
\begin{equation}
\label{f8} 
\Omega_2=\frac{<t_1G_0t_2G_0t_1+t_2G_0t_1G_0t_2 >}{<T^0_{K D}>}
\end{equation}
and
\begin{equation}
\label{f9} 
 \Sigma_2=\frac{<T^0_{K D}G_{NN}T^0_{K D}G_{NN}T^0_{K D}>}
{<T^0_{K D}>} .
\end{equation}
\begin{equation}
\label{f9a} 
\Delta_2=2\frac{<T^0_{K D}G_{NN}(t_1G_0t_2+t_2G_0t_1)>}{<T^0_{K D}>}
\end{equation}

This procedure may be continued but the main advantage of this 
order is a strong 
cancellation in the $\Sigma_2 - \Sigma_1^2$ term (and indeed also 
in the higher order $\Sigma_n$ terms).

The convergence of our expansion is shown in Table 1 in the case 
of $S$ wave scattering length. It seems quite satisfactory 
and  a  percent precision is reached.
The details of the calculations may be found in Ref. \cite{etad}
where this formalism is used to describe the $\eta$ deuteron scattering.

For completeness, the scattering length should contain 
also an $ inner $ Coulomb correction. A quick estimate follows 
from a boundary condition model that  generates a relation   
$1/A_{c} = 1/A - 2/B (ln(2R_D/B)+1.154)$ where $A_c$ is the 
Coulomb corrected length. The change is on a percent level 
and will be neglected in the following estimates.

\section{RESULTS}

The  meson-deuteron scattering length
$A_{K D}$ is expressed in terms of  "effective
$Kp$ and $Kn$ scattering lengths" $\bar a_{K N}$, which are some
averages of the $K N$ scattering matrices extrapolated
below the threshold. These are input parameters  taken from 
several analyses of the  combined $\pi \sigma, \pi \Lambda,K N $ 
coupled channels data. A collection  of characteristic 
numbers is given in Table 2.

In Table 3 one finds values of the atomic 
level shifts and half-widths  for the $K N$ 
effective scattering lengths given in Table 2. 
The resulting $S$ wave  length 
$A_{K D}$ and related level width may vary by an order of
magnitude, reflecting the nearby $K D$ quasibound
state that may arise at threshold for $\Re a_{K N}$
close to 1 fm. It is clear that such a possibility 
exists for one of the models discussed \cite{tana},\cite{schn}. 
Other models are rather distant from this situation.  

The effective lengths were grouped into two sets. Those of 
Refs. \cite{chao},\cite{amar},\cite{kais},\cite{land} 
fit the scattering results and disagree with the $1S$ kaonic
hydrogen level data. The differences of the $KD$ level shifts 
and widths are rather moderate, still some of those, 
in particular the $2P$ level widths could be discernible 
by the atomic cascade measurements. The differences involved 
in these models are partly related to the subthreshold 
extrapolations which are known to depend on details of 
the interaction. Other differences are due to different 
nonresonant $ I =1 $ amplitudes that weight heavily in the 
deuteron case.

Other, less conventional, models tend to describe the 
hydrogen level shift and width. Those 
\cite{tana},\cite{kuma},\cite{schn} generate significantly 
smaller $KN$ absorptive amplitudes. As a  consequence the 
widths of  $2P,3D$ states in the deuterium are also much smaller. 
Another interesting physical consequence of some of these 
models is a near-possibility 
of a nuclear  $S$ wave $KD$ state. Such a state is 
almost-generated by the attractive $Kn$ interactions involved 
in  models of Refs. \cite{tana},\cite{schn}. Proximity of 
such a state is reflected in the enhanced widths of the $1S$ 
kaonic-deuterium atomic level indicated in Table 3.

Two conclusions follow from this paper:

\noindent a) Different parametrisations and models of 
kaon-nucleon interactions generate sizable differences 
in the kaonic-deuteron level shifts and lifetimes.

\noindent b) Two groups of models,  that reproduce 
either the scattering or the hydrogen data,  
generate very different $2P,3D$ deuteron widths. 
These two situations should be discernible by the 
atomic cascade in the kaonic deuterium even if the shape 
of $2P \rightarrow 1S$ line turns out difficult to resolve.

\begin{table}
\caption{Convergence of the $A_{K D}(L=0)$ expansion. Last line 
is a third order formula, not included in the text.
$ \bar{a}_{K n}=-0.1 + i0.5, \bar{a}_{K p} = -1. + i1.5 $  }
\begin{tabular}{lcl}
$A_{K D}^0 $ & -1.39 + i2.42 &  $Eq.(\ref{f11})$ \\
$A_{K D}^1 $ & -1.38 + i0.74 &  $Eq.(\ref{f5})$ \\
$A_{K D}^2 $ & -1.23 + i0.95 &  $Eq.(\ref{f7})$ \\
$A_{K D}^3 $ & -1.24 + i0.95 &  $Order 3$ \\
\end{tabular}
\label{table1}
\end{table}

\begin{table}
\caption{
Kaon-nucleon "effective lengths" in $fm$ units. These numbers 
have been calculated from figures in the referred papers, hence 
contain some numerical uncertainties and have been rounded 
correspondingly.}
\begin{tabular}{lcc}
               	       &    $\bar{a_{K p}}$   & $\bar{a_{K n}}$ \\
Martin \cite{amar}       & -1.0 + i1.5  & -0.1 + i0.5   \\
Chao  \cite{chao}        & -0.7 + i1.8  &  0.4 + i0.7   \\
Ha [3] \cite{land}       & -1.0 + i2.0  & -0.1 + i0.7   \\
Waas \cite{kais}         & -0.5 + i2.2  &  0.8 + i0.5   \\
Tanaka \cite{tana}       &  0.0 + i0.7  &  1.1 + i0.5   \\
Kumar \cite{kuma}        & -1.0 + i0.1  &  0.4 + i0.7   \\
Schnick \cite{schn},\cite{tana} & 0.6 + i0.5 & 1.1 + i0.2\\
\end{tabular}
\label{table2}
\end{table}

\begin{table}
\caption{ Level shifts and half-widths of the $1S,2P,3D$ 
states in kaonic-deuterium. The units are 
 $ KeV,meV,10^{-2}\mu eV$ correspondingly.}

\begin{tabular}{lccc}
   & $ \Delta E-i\Gamma/2_{1S}$ & $ \Delta E-i\Gamma/2_{2P}$ & 
$ -i\Gamma/2_{3D}$  \\
Martin \cite{amar}       & 0.76 - i0.64  & 6.5 - i12  &  - i6.9  \\
Chao  \cite{chao}        & 0.74 - i0.78  & 1.4 - i15  &  - i8.5  \\
Ha [3] \cite{land}       &-0.81 - i0.75  & 5.7 - i16  &  - i9.3  \\
Waas \cite{kais}         & 0.83 - i0.87  &-2.5 - i16  &  - i9.4  \\
Tanaka \cite{tana}       & 0.29 - i1.22  &-6.7 - i6.9 &  - i4.2  \\
Kumar \cite{kuma}        & 0.43 - i0.19  & 3.8 - i4.0 &  - i2.3  \\
Schnick \cite{schn},\cite{tana} & 0.58 - i2.87 & -9.8 -i3.7 & - i2.4  \\
\end{tabular}
\label{table3}
\end{table}

\end{document}